\documentclass[9pt,article,twoside]{rmaa-rho-class/rmaa-rho}
\RMxAAtemplatetype{\RMxAA} 

\usepackage{url}
\usepackage{comment}

\vol{61}
\pages{1-6}
\thisyear{2026}
\doi{\href{https://www.astroscu.unam.mx/rmaa/RMxAA..XX-X}{https://www.astroscu.unam.mx/rmaa/RMxAA..XX-X}}

\title{The Space Debris Tracking and Surveillance program of the IAC80 telescope}

\author[1,2$\dagger$]{Olga Zamora \orcidlink{0000-0003-2100-1638}}

\affil[1]{Instituto de Astrofísica de Canarias, E-38205, La Laguna,  Tenerife, Spain}
\affil[2]{Universidad de La Laguna, Departamento de Astrofísica, E-38206, La Laguna, Tenerife, Spain}

\leadauthor{Olga Zamora}
\smalltitle{ }

\corres{Olga Zamora}
\email{olga.zamora@iac.es}

\received{xxx}
\accepted{\today}

\license{Texto de la licencia aquí}

\setbool{rho-abstract}{true} 
\setbool{rho-resumen}{true} 

\begin{abstract}
The increasing population of artificial objects in Earth orbit poses significant risks to operational spacecraft and ground infrastructure, making Space Surveillance and Tracking (SST) a critical activity. Since 2015, the IAC80 telescope at the Teide Observatory has been contributing to European SST efforts through a dedicated space debris tracking program. Originally designed for astronomical observations, the IAC80 has been progressively adapted to perform automated monitoring of satellites and debris objects in Low Earth Orbit (LEO), Medium Earth Orbit (MEO), and Geostationary Earth Orbit (GEO).

This work presents an overview of the IAC80 SST system, including the telescope and site characteristics, the optical instruments currently in use, and the automation tools developed to enable autonomous operations. Particular emphasis is placed on the real-time data reduction pipeline, SATRED, which performs calibration, automatic satellite trail detection, and astrometric measurements during observations. The trail detection algorithm combines robust background estimation, contour extraction, and geometric characterization to provide reliable positional and photometric information.

The IAC80 system routinely delivers accurate measurements, with astrometric precision better than one arcsecond and the capability to acquire a large number of observations per night, particularly for MEO and GEO targets. These results demonstrate the effectiveness of adapting classical astronomical facilities for operational SST applications and highlight the role of the IAC80 telescope as a reliable contributor to the European Space Surveillance and Tracking network.
\end{abstract}

\keywords{keyword 1, keyword 2, keyword 3, keyword 4, keyword 5}

\begin{resumen}
El creciente número de objetos artificiales en órbita terrestre supone riesgos significativos para las misiones espaciales operativas y para las infraestructuras en tierra, lo que convierte la Vigilancia y Seguimiento Espacial (Space Surveillance and Tracking, SST) en una actividad crítica. Desde 2015, el telescopio IAC80 del Observatorio del Teide contribuye a los esfuerzos europeos de SST mediante un programa dedicado de seguimiento de basura espacial. Diseñado originalmente para observaciones astronómicas, el IAC80 ha sido adaptado progresivamente para realizar el monitoreo automático de satélites y objetos de desecho en órbitas bajas (LEO), medias (MEO) y geoestacionarias (GEO).

Este trabajo presenta una visión general del sistema SST del IAC80, incluyendo las características del telescopio y del emplazamiento, los instrumentos ópticos actualmente en uso y las herramientas de automatización desarrolladas para permitir operaciones autónomas. Se pone especial énfasis en el pipeline de reducción de datos en tiempo real, SATRED, que realiza la calibración, la detección automática de trazas de satélites y las mediciones astrométricas durante las observaciones. El algoritmo de detección de trazas combina una estimación robusta del fondo, la extracción de contornos y la caracterización geométrica para proporcionar información posicional y fotométrica fiable.

El sistema IAC80 proporciona de forma rutinaria mediciones precisas, con una precisión astrométrica mejor que un segundo de arco y la capacidad de adquirir un gran número de observaciones por noche, especialmente para objetivos en MEO y GEO. Estos resultados demuestran la eficacia de adaptar instalaciones astronómicas clásicas para aplicaciones operativas de SST y destacan el papel del telescopio IAC80 como un contribuidor fiable a la red europea de Vigilancia y Seguimiento Espacial.
\end{resumen}

\begin{document}
\nolinenumbers

\maketitle
\pagestyle{fancy}\thispagestyle{firststyle}


\section{INTRODUCTION}
The rapid growth of space activities over the last decades has resulted in a significant increase in the number of satellites and debris objects orbiting the Earth. This situation poses risks to active missions due to potential collisions and uncontrolled re-entries. Space Surveillance and Tracking (SST) systems are therefore essential to monitor the orbital environment, characterize space objects, and provide timely information to protect both people and space-based and ground-based assets.

Optical ground-based telescopes play a fundamental role in SST, particularly for objects in Medium Earth Orbit (MEO) and Geostationary Earth Orbit (GEO). In this context, the Instituto de Astrofísica de Canarias (IAC) initiated in 2015 a space debris tracking and surveillance program using the IAC80 telescope at the Teide Observatory. The objective was to adapt an existing astronomical facility into an automated and efficient SST sensor capable of regular operations and integration within European SST activities.

This paper presents an overview of the IAC80 SST program, including the telescope and site characteristics, the instruments currently in use, the automation tools developed to enable autonomous observations, and the real-time processing pipeline designed for satellite trail detection and astrometric measurements.

\section{The IAC80 telescope and Observing site}

The IAC80 telescope is located at the Teide Observatory (OT), on the island of Tenerife, Spain. The observatory benefits from excellent atmospheric conditions, with low cloud coverage, stable seeing, and more than 300 useful observing nights per year. These characteristics make the site particularly suitable for continuous and long-term SST operations.

The IAC80 is a classical Cassegrain telescope with an aperture of 82 cm and a focal length of approximately 9 m, corresponding to a focal ratio of f/11.3. The telescope was inaugurated in 1995 and it was the first telescope designed and built by Spain. Although originally designed for astronomical observations, its mechanical stability and pointing capabilities make it well suited for satellite tracking tasks.

Since 2015, the IAC80 has been routinely used to observe objects in different orbital regimes, including Low Earth Orbit (LEO), MEO, and GEO. The telescope operates in coordination with dedicated software tools that allow the execution of fully automated observation sequences during SST campaigns.

\section{Instrumentation: CAMELOT2 and CARONTE}
\subsection{CAMELOT2}
CAMELOT2 \citep{camelot2} is the main common-user instrument of the IAC80 telescope. It replaced the previous CAMELOT camera in mid-2019. The name CAMELOT2 comes from the Spanish Cámara Mejorada Ligera del Observatorio del Teide (Improved Lightweight Camera of the Teide Observatory).

The instrument is equipped with a 4k × 4k back-illuminated CCD (E2V Technologies CCD231-84) manufactured by Spectral Instruments (1100 SI Series Camera). The detector operates in the optical wavelength range and has a pixel size of 15 µm, corresponding to an on-sky scale of 0.326 arcseconds per pixel. The theoretical field of view is 22 × 22 arcminutes²; however, due to vignetting caused by the filters, the useful square field of view is approximately 11.8 × 11.8 arcminutes².

CAMELOT2 is complemented by an extensive set of filters, including SDSS griz, Johnson–Cousins UBVRI, Strömgren uvby, and several narrow-band filters. This versatility allows the instrument to be used both for classical astronomical programs and for SST observations, including satellite photometry and characterization.

\begin{figure}[ht]
            \centering
            \includegraphics[width=0.99\columnwidth]{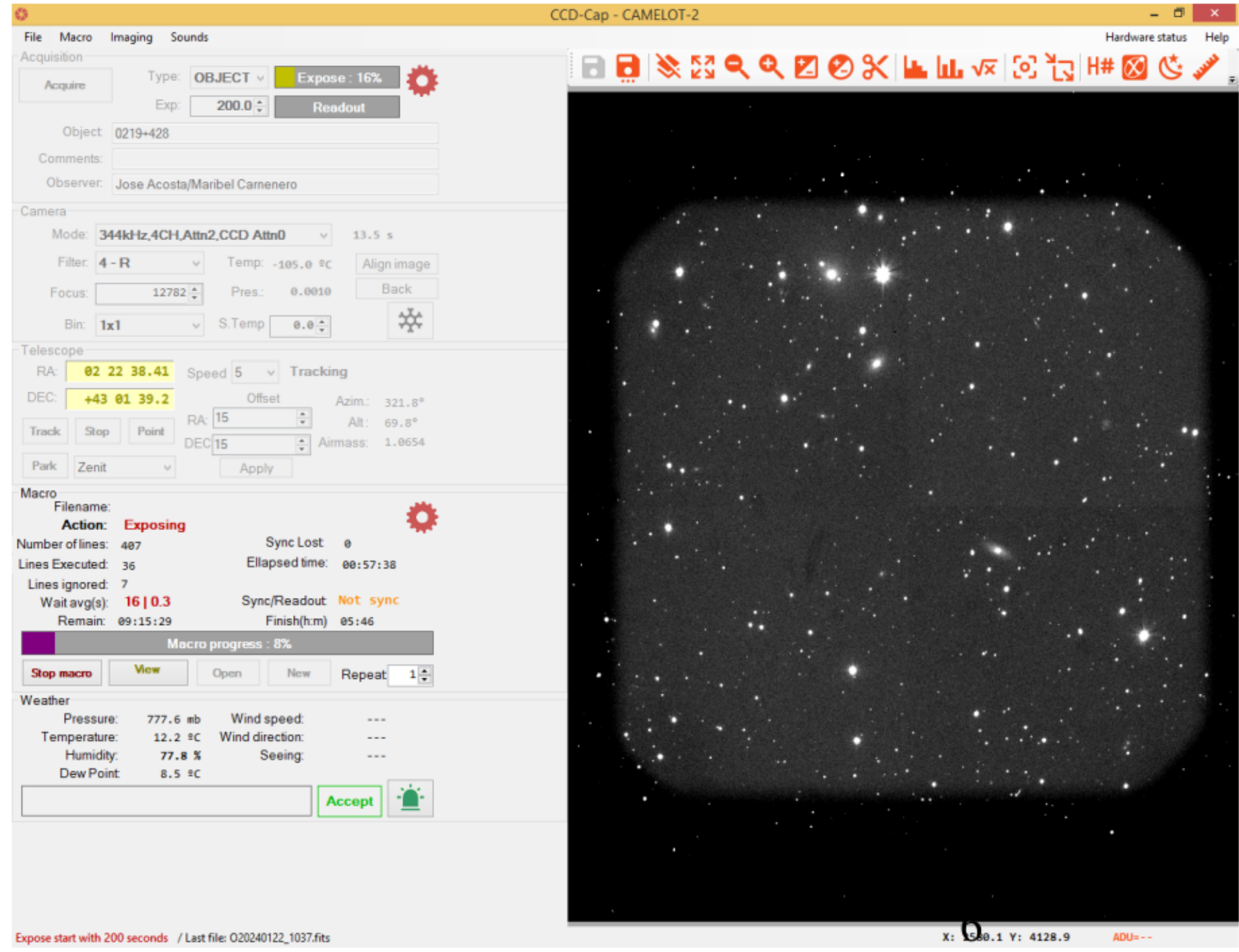}
            \caption{Graphical user interface of the CCD-Cap control interface used with the CAMELOT2 instrument at the IAC80 telescope during an observation. The graphical user interface provides real-time monitoring of telescope pointing, camera configuration, exposure status, macro execution, and displays the acquired CCD image.}
            \label{ccdcap}
        \end{figure}

Figure \ref{ccdcap} illustrates the CCD-Cap control software used to operate the CAMELOT2 instrument at the IAC80 telescope. The graphical user interface (GUI), developed in C\#, provides integrated control of the camera, telescope, and observing macros during SST operations. It displays in real time the telescope pointing coordinates, camera configuration parameters, filter selection, exposure and readout status, as well as the execution progress of automated observing sequences. The interface also shows the acquired CCD image immediately after readout, allowing on-site monitoring of image quality and target detection. This software environment plays a key role in enabling fully automated observations with CAMELOT2, supporting efficient and reliable space debris tracking during routine operations.

\subsection{CARONTE}
In addition to CAMELOT2, since 2022 the IAC80 is also equipped with the CARONTE camera (Cámara de Alta resolución súper Rápida del Observatorio Nocturno del TEide, translated into English as Super-fast high-resolution camera of the night Teide Observatory \citet{caronte}), a modern sCMOS-based instrument. CARONTE uses a Sony IMX461 medium-format sensor with 102 megapixels and a 16-bit analog-to-digital converter. The pixel size is 3.76 µm, resulting in an angular scale of 0.081 arcseconds per pixel and a field of view of 15.9 × 12 arcminutes².

CARONTE is controlled via a Python-based graphical user interface and is designed for fast imaging applications. Its high spatial resolution and low readout noise make it particularly suitable for detailed characterization of satellites and space debris, especially due to its small read-out time.

\begin{figure}[ht]
            \centering
            \includegraphics[width=0.99\columnwidth]{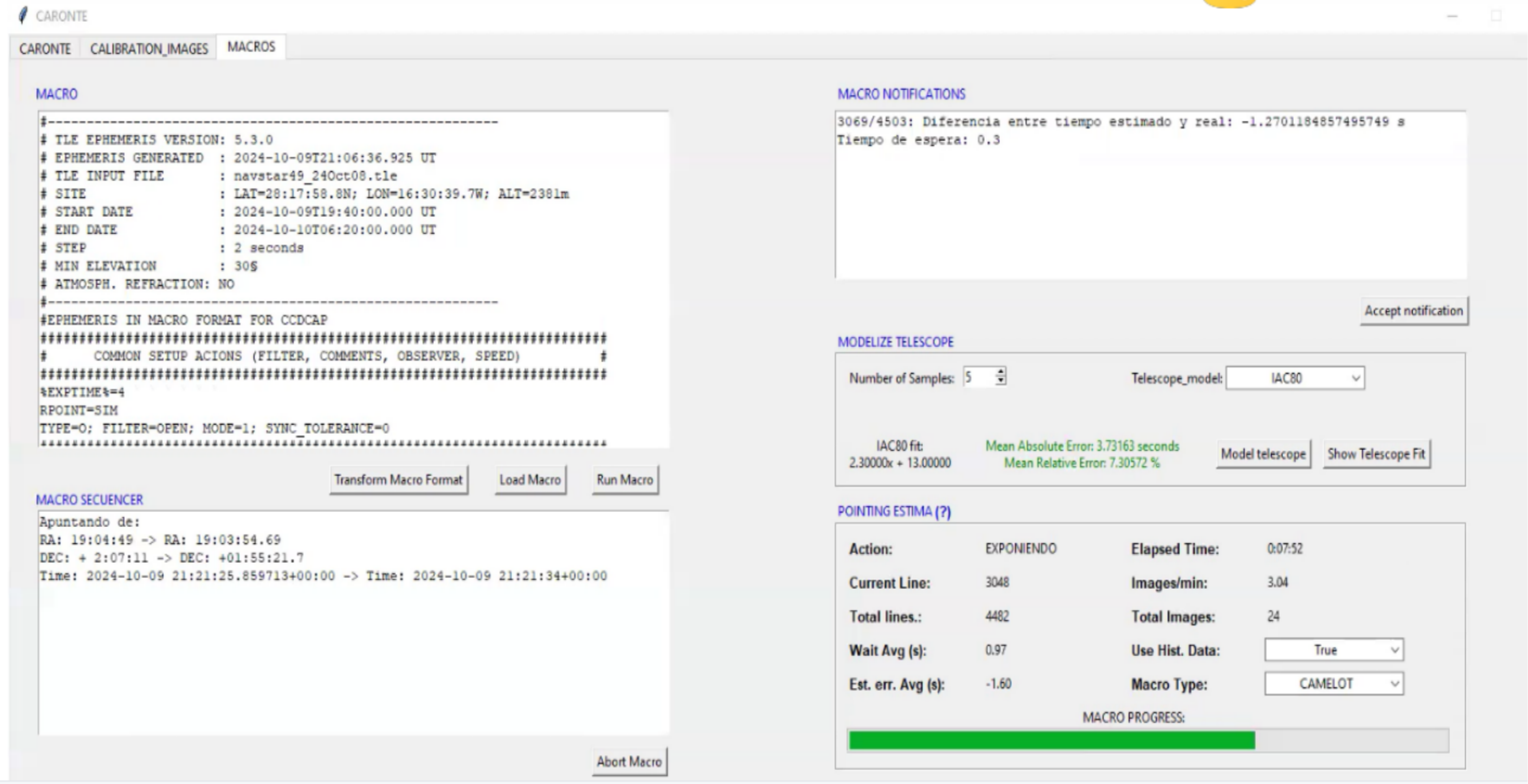}
            \caption{Graphical user interface of the CARONTE control software used during SST operations with the IAC80 telescope. The interface displays the ephemeris-based macro definition, macro sequencing, pointing estimation, telescope model fitting, and real-time monitoring of exposure progress and acquisition status.}
            \label{caronte}
        \end{figure}

Figure \ref{caronte} shows the GUI developed to control the CARONTE instrument during SST observations with the IAC80 telescope. The software provides an integrated environment for the definition and execution of ephemeris-based observing macros, allowing the automatic sequencing of pointings and exposures. The interface includes tools for pointing estimation and telescope model fitting, which are used to refine the telescope pointing accuracy during the observing session. Real-time information on exposure progress, acquisition status, and macro execution is displayed, enabling continuous monitoring of the observation flow. This control software is a key element in the automation of CARONTE operations, supporting efficient and reliable data acquisition for the tracking and characterization of satellites and space debris.

\section{Automation and Control Software}
To enable efficient SST operations, several software tools have been developed for the IAC80 telescope. These tools allow almost fully automated observations with minimal human intervention.

A dedicated graphical user interface (GUI) manages the execution of observation macros, instrument configuration, and data acquisition. An Ephemeris Selector module is used to plan observations based on predicted satellite positions, visibility constraints, and orbital priorities.

A key component of the system is the Pointing Estimator, which implements a self-learning approach. By analyzing the results of previous pointings, the estimator improves the telescope pointing accuracy over time. It calculates the time needed to reach a desired position from an initial position to a final position in right ascension and declination, based on a table of mean time and standard deviation obtained from pointings.
This adaptive behavior is particularly important for tracking fast-moving or faint objects and contributes to the overall reliability of the system.

\begin{figure}[ht]
            \centering
            \includegraphics[width=0.99\columnwidth]{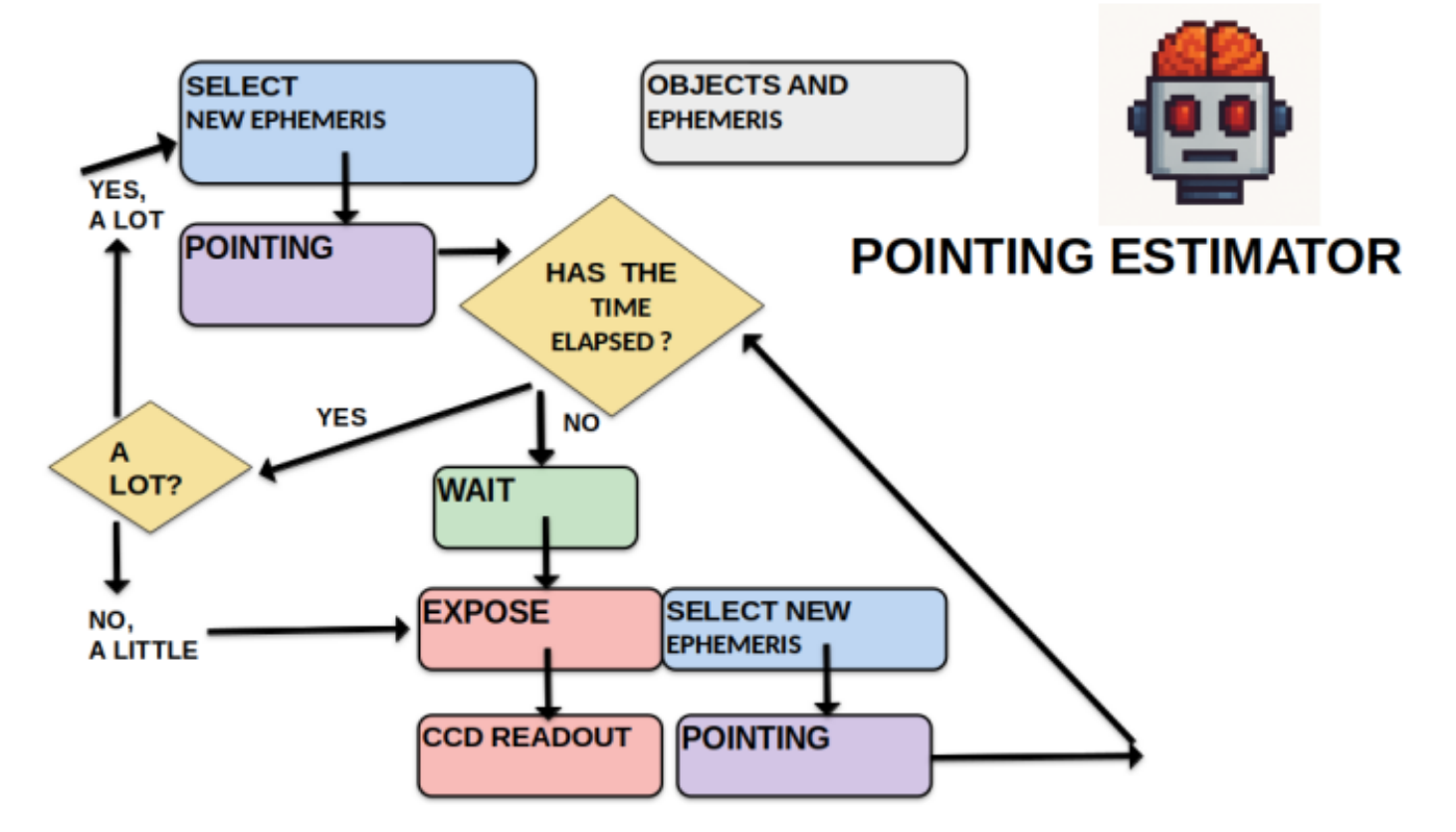}
            \caption{Operational workflow of the CAMELOT2 ephemeris selector and pointing estimator. The diagram illustrates the automated loop for ephemeris selection, telescope pointing, timing control, exposure execution, and adaptive pointing correction based on feedback from previous observations.}
            \label{estimator}
        \end{figure}

Figure \ref{estimator} illustrates the operational logic implemented in the CAMELOT2 Ephemeris Selector and its interaction with the Pointing Estimator, which together control the automated observation loop of the IAC80 telescope during Space Surveillance and Tracking sessions. This logic is also used for the CARONTE Pointing Estimator.

The workflow starts with the selection of a new ephemeris, corresponding to a specific target object and its predicted position as a function of time. The ephemeris information is retrieved from the object database and passed to the pointing module. Based on this information, the telescope performs a pointing operation, slewing to the expected right ascension and declination of the target.

Once the pointing is completed, the system evaluates whether the scheduled observation time has elapsed. If the required time has not yet been reached, the system enters a waiting state, ensuring synchronization between the ephemeris prediction and the actual observation time. When the appropriate time window is reached, an exposure is executed, followed by the CCD readout.

After the image acquisition, the system assesses whether the pointing error observed in previous exposures is significant (“a lot”) or small (“a little”). If a large deviation is detected, the workflow loops back to the selection of a new ephemeris, triggering an updated pointing command to correct the telescope position. If the deviation is small, the system proceeds directly with the exposure sequence, optimizing observing efficiency by avoiding unnecessary re-pointing.

The Pointing Estimator operates as a feedback component throughout the loop. It continuously incorporates information from previous pointings and observations to refine the expected telescope offsets. This adaptive behavior allows the system to progressively improve pointing accuracy over time, reducing systematic errors and increasing the success rate of automatic observations.

Overall, this workflow enables autonomous operation of the IAC80 during SST campaigns, combining ephemeris-driven scheduling, real-time decision making, and self-learning pointing correction to efficiently track satellites and space debris objects.
Together, these tools allow the system to follow predefined schedules and adapt observations as needed.

\section{Real-Time Data Reduction Pipeline: SATRED}
The data acquired during SST observations are processed by a real-time pipeline called SATRED (SATellite REDuction). The pipeline is designed to provide rapid feedback during observations and to generate astrometric and photometric measurements on-the-fly automatically.

\subsection{Operational workflow}
SATRED performs the standard calibration steps (bias subtraction and flat-field calibration) and then applies an automatic satellite trail detection algorithm. The background level and noise are estimated using robust statistics, and pixels above a given threshold are selected. A mild Gaussian smoothing is applied to merge fragmented trail segments, after which contours are extracted.

 \begin{figure}[ht]
            \centering
            \includegraphics[width=0.99\columnwidth]{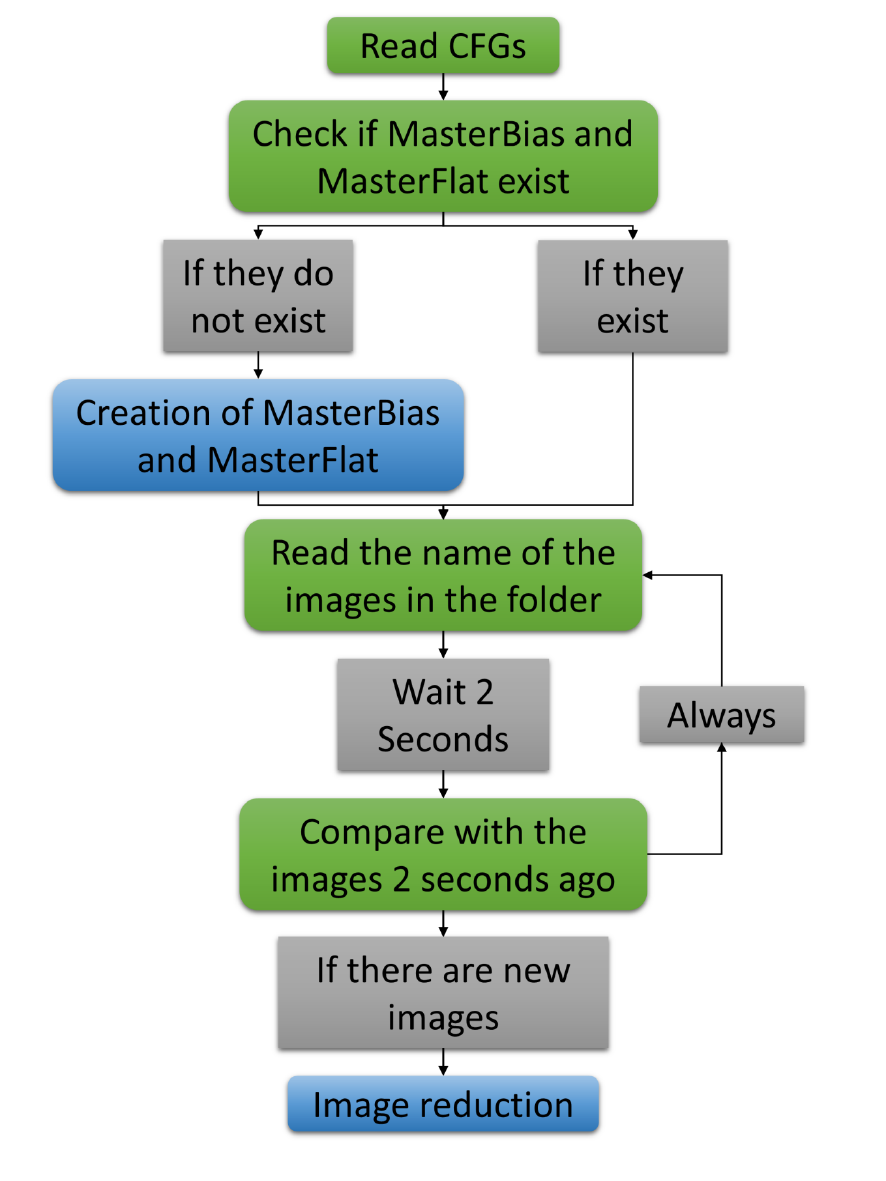}
            \caption{Global workflow of the SATRED pipeline. The diagram shows the initialization stage, including configuration reading and verification or creation of master calibration frames, followed by continuous monitoring of the image directory to detect newly acquired images and automatically trigger the reduction process.}
            \label{satred_1}
        \end{figure}

\begin{figure}[ht]
            \centering
            \includegraphics[width=0.99\columnwidth]{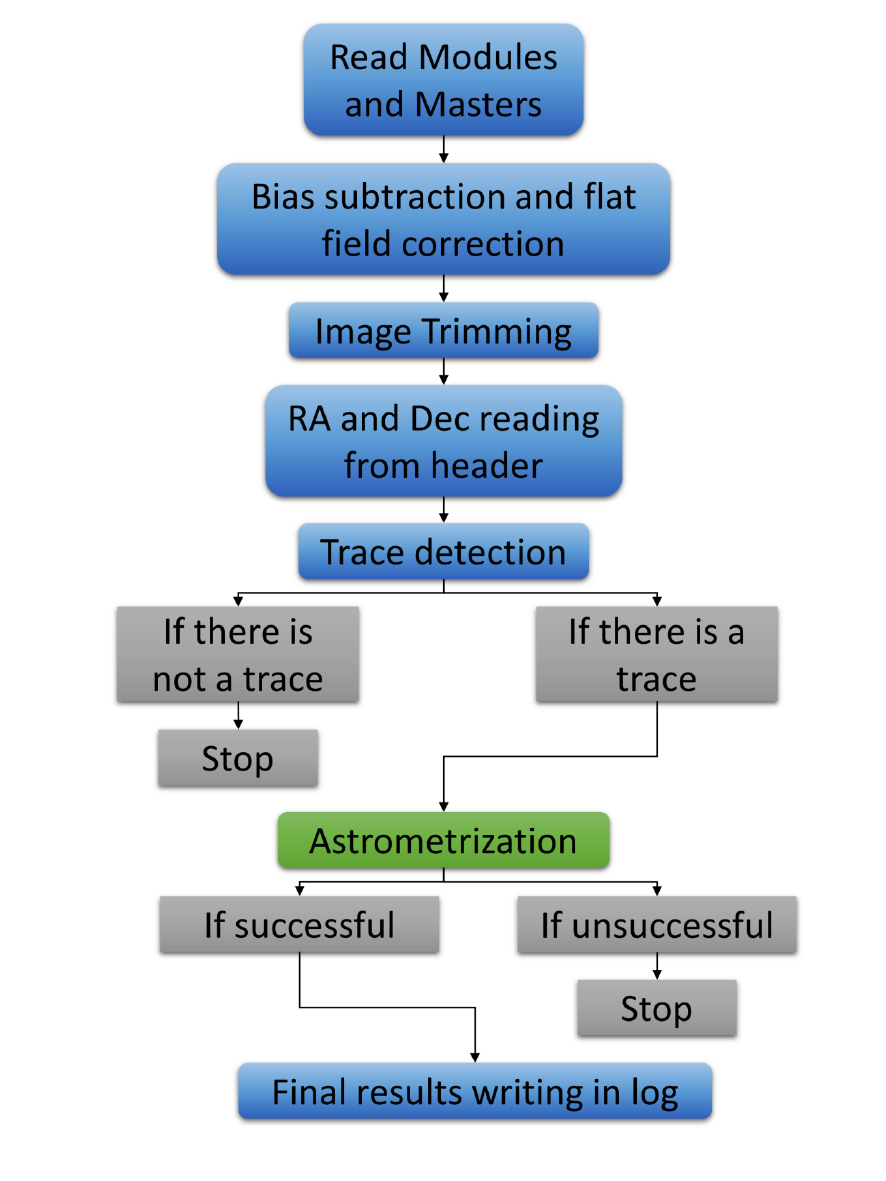}
            \caption{Internal image reduction and satellite trail detection workflow implemented in SATRED. The process includes bias subtraction and flat-field correction, image trimming, extraction of pointing information from the FITS header, automatic trace detection, astrometrization of detected trails, and logging of final results.}
            \label{satred_2}
        \end{figure}

 Figure \ref{satred_1} illustrates the global control logic of SATRED. The pipeline starts by reading the configuration files, which define the instrument parameters, directory structure, and processing options. At initialization, SATRED checks for the existence of the corresponding master calibration frames, namely the master bias and master flat. If these calibration products are not available, they are automatically generated from the appropriate calibration images before continuing with the reduction process.

Once the calibration frames are available, SATRED continuously monitors the image directory. The pipeline reads the list of image filenames in the folder and compares it with the state from a few seconds earlier, using a fixed time interval of two seconds. This mechanism allows SATRED to identify newly acquired images as soon as they appear. When new images are detected, the pipeline automatically triggers the image reduction process.

The internal image reduction workflow is detailed in Figure \ref{satred_2}. For each new image, the required modules and master calibration frames are loaded. The image is first corrected for instrumental effects through bias subtraction and flat-field correction. A trimming step is then applied to remove non-illuminated regions or overscan areas. After calibration, the right ascension and declination of the telescope pointing are read directly from the FITS header and associated with the image.

The next stage consists of the automatic trace detection. The algorithm analyzes the calibrated image to identify elongated features consistent with satellite or debris trails. If no trace is detected, the pipeline stops processing the image. When a trace is found, SATRED proceeds with the astrometrization stage, where the trail position is measured in celestial coordinates. If the astrometric solution is successful, the final results — including positional information and quality indicators — are written to a log file for further analysis. If astrometrization fails, the image is discarded and the pipeline continues with the next acquisition.

This modular and event-driven design allows SATRED to operate continuously during observing sessions, providing rapid feedback and supporting autonomous SST operations. The pipeline plays a central role in enabling efficient, reliable, and scalable satellite tracking with the IAC80 telescope.

\subsection{Satellite Trail Detection Algorithm}

The satellite trail detection algorithm implemented in SATRED is designed to identify elongated structures associated with satellite or space debris motion in calibrated CCD images. The algorithm operates on a single image and combines background estimation, statistical thresholding, contour analysis, and geometric filtering.

\subsubsection{Background Estimation and Image Thresholding}

Let $I(x,y)$ be the calibrated image. The global background level is estimated as the median of the image:

\begin{equation}
B = \mathrm{median}\left( I(x,y) \right)
\end{equation}

An initial estimate of the global dispersion is computed as the standard deviation of the full image, $\sigma_{\mathrm{all}}$. To obtain a more robust estimate of the background noise, a second dispersion estimator is computed using only pixels below a background-dependent threshold:

\begin{equation}
\sigma_{\mathrm{back}} = \mathrm{std}\left( I(x,y) \;\middle|\; I(x,y) < B + 0.5\,\sigma_{\mathrm{all}} \right)
\end{equation}

Pixels with negative values are replaced by the background level $B$ to avoid spurious detections.

Candidate trace pixels are selected by applying a threshold based on the background dispersion:

\begin{equation}
M(x,y) =
\begin{cases}
1, & \text{if } I(x,y) > B + N_{\sigma}\,\sigma_{\mathrm{back}} \\
0, & \text{otherwise}
\end{cases}
\end{equation}

where $N_{\sigma}$ corresponds to the \texttt{nsigmas} parameter of the algorithm.

The resulting binary mask is smoothed using a Gaussian filter with a $7 \times 7$ kernel in order to merge fragmented segments of the same trace and suppress isolated noise peaks.

\subsubsection{Contour Extraction and Geometric Selection}

Contours are extracted from the smoothed binary image using an external contour retrieval method. For each contour, a convex hull is computed, and its area $A_i$ is evaluated. Contours with $A_i < A_{\mathrm{min}}$ or with fewer than five vertices are discarded.

For the remaining candidates, a minimum-area bounding rectangle is fitted to the convex hull. The rectangle is characterized by its center $(x_i,y_i)$, its sides $(w_i,h_i)$, and its position angle $\theta_i$. The longer and shorter sides of the rectangle are defined as:

\begin{equation}
L_i = \max(w_i,h_i), \qquad S_i = \min(w_i,h_i)
\end{equation}

An elongation ratio is computed as:

\begin{equation}
R_i = \frac{L_i}{S_i}
\end{equation}

Only candidates with $R_i > R_{\mathrm{min}}$ are retained as valid satellite trail detections, where $R_{\mathrm{min}}$ corresponds to the \texttt{min\_ratio} parameter.

The centroid of each valid trace is computed from the first-order moments of the contour.

\subsubsection{Flux Measurement and Local Background Subtraction}

For each detected trace, a rectangular mask aligned with the fitted bounding rectangle is generated and used to define the trace aperture. A second rectangular mask, expanded by a configurable separation parameter $d$ (\texttt{rect\_sep}), is constructed to isolate the trace from the surrounding background.

A larger rectangular sky region is then defined around the trace aperture, excluding the separation region. The local background level is estimated as the median pixel value within this sky region:

\begin{equation}
B_{\mathrm{local}} = \mathrm{median}\left( I(x,y) \in \mathcal{S} \right)
\end{equation}

The integrated flux associated with the trace is computed as:

\begin{equation}
F_{\mathrm{trace}} = \sum_{(x,y) \in \mathcal{T}} I(x,y)
\end{equation}

and the background contribution is removed according to:

\begin{equation}
F_{\mathrm{net}} = F_{\mathrm{trace}} - B_{\mathrm{local}} \cdot N_{\mathcal{T}}
\end{equation}

where $N_{\mathcal{T}}$ is the number of pixels within the trace aperture.

\subsubsection{Post-processing and Output Parameters}

If multiple trace segments are detected in a single image, their properties are reported as separated detections by providing their individual centroid positions and orientation angles, and summing areas, lengths, and integrated fluxes.

The final output of the algorithm consists of the trace centroid coordinates, total area in pixels, integrated flux, trace length, and the parameters of the fitted bounding rectangle. These quantities are subsequently passed to the astrometric processing stage of the SATRED pipeline.

Table~\ref{tab:trail_params} summarizes the main configurable parameters of the satellite trail detection algorithm implemented in SATRED. These parameters control the sensitivity of the detection, the geometric selection of elongated structures, and the definition of the photometric regions used for flux measurement. The detection threshold, expressed in units of the background dispersion, determines the balance between sensitivity to faint trails and robustness against noise. The minimum area and elongation ratio parameters are used to reject spurious detections and compact sources, ensuring that only features consistent with satellite or debris traces are retained. The separation parameter defining the distance between the trace aperture and the surrounding sky region plays a key role in obtaining a reliable local background estimate. Together, these parameters allow the algorithm to be tuned to different observing conditions and orbital regimes while maintaining stable and repeatable performance.

\begin{figure}[ht]
            \centering
            \includegraphics[width=0.99\columnwidth]{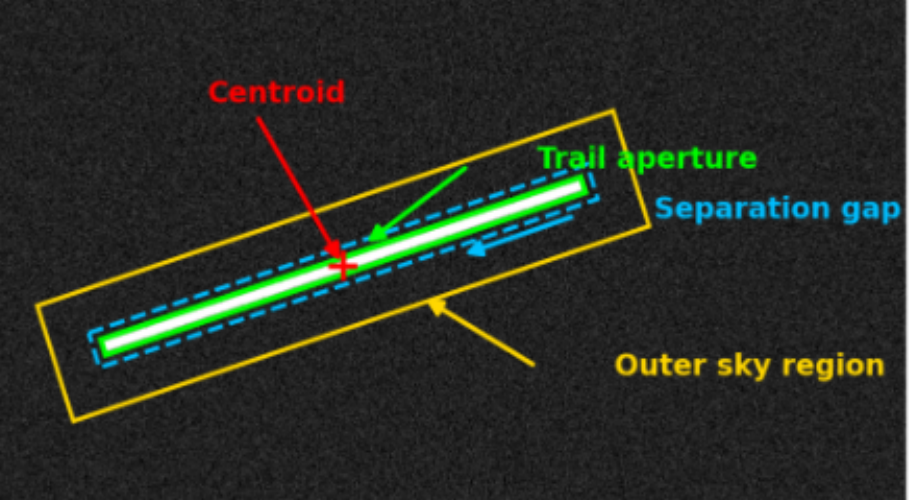}
            \caption{Schematic representation of the satellite trail photometry and background estimation scheme used in the SATRED detection algorithm. The trail aperture is aligned with the detected trace orientation, while a surrounding rectangular sky region, separated by a configurable gap, is used to estimate the local background.}
            \label{trail}
        \end{figure}

Finally, Figure \ref{trail} illustrates the geometric definition of the photometric regions used by the SATRED trail detection algorithm once a satellite trace has been identified. The centroid of the trace is computed from the contour moments and is indicated at the center of the detected structure. A rectangular aperture, aligned with the orientation of the fitted minimum-area bounding rectangle, is used to integrate the signal associated with the satellite trail.

To obtain a reliable estimate of the local sky background, an outer rectangular region is defined around the trail aperture. This region is separated from the trace by a configurable gap in order to avoid contamination from the trail signal itself. The background level is computed as the median pixel value within this outer sky region and is subsequently subtracted from the integrated flux measured inside the trail aperture. This geometry allows accurate flux estimation even in the presence of background gradients or non-uniform sky conditions.
The algorithm has been extensively tested under different observing conditions and orbital regimes, providing stable performance for routine SST operations.

\begin{table*}[t]
\centering
\caption{Main parameters of the satellite trail detection algorithm implemented in SATRED.}
\label{tab:trail_params}
\begin{tabular}{l c p{9cm}}
\hline
Parameter & Symbol & Description \\
\hline
\texttt{nsigmas} & $N_{\sigma}$ &
Detection threshold expressed as the number of background standard deviations above the median sky level used to select candidate trace pixels. \\

\texttt{min\_area} & $A_{\mathrm{min}}$ &
Minimum area, in pixels, of the convex hull associated with a detected contour. Candidates with smaller areas are rejected as noise or spurious detections. \\

\texttt{min\_ratio} & $R_{\mathrm{min}}$ &
Minimum elongation ratio between the major and minor axes of the fitted minimum-area bounding rectangle. This parameter enforces the selection of elongated structures consistent with satellite trails. \\

\texttt{rect\_sep} & $d$ &
Separation, in pixels, between the trace aperture and the surrounding background region used for local sky estimation during flux measurement. \\

Gaussian kernel size & -- &
Size of the Gaussian smoothing kernel applied to the thresholded image to merge fragmented trace segments and suppress isolated noise peaks. \\

Background cutoff factor & $k$ &
Multiplicative factor applied to the global dispersion estimate to select background-dominated pixels for robust noise estimation. \\

\hline
\end{tabular}
\end{table*}

\section{Performance and Results}
\begin{figure*}[t]
\centering
\includegraphics[width=0.32\textwidth]{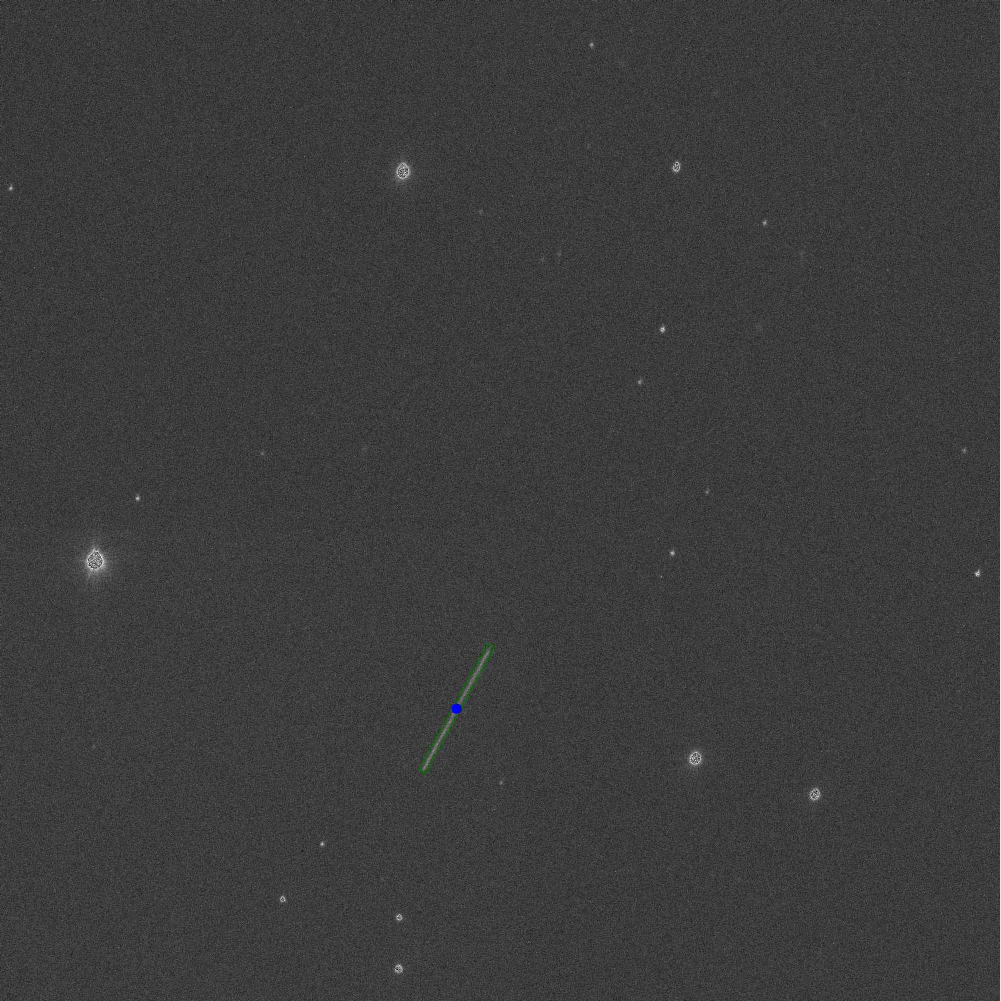}\hfill
\includegraphics[width=0.32\textwidth]{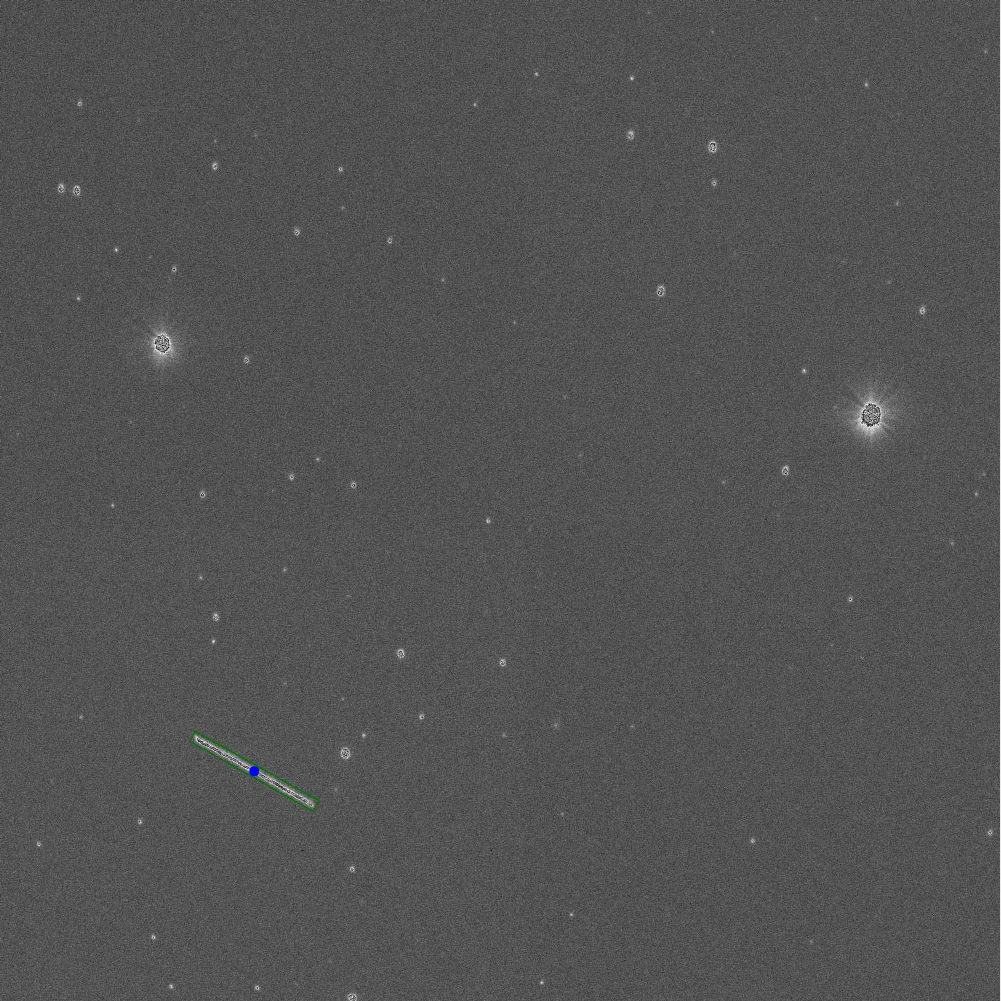}\hfill
\includegraphics[width=0.32\textwidth]{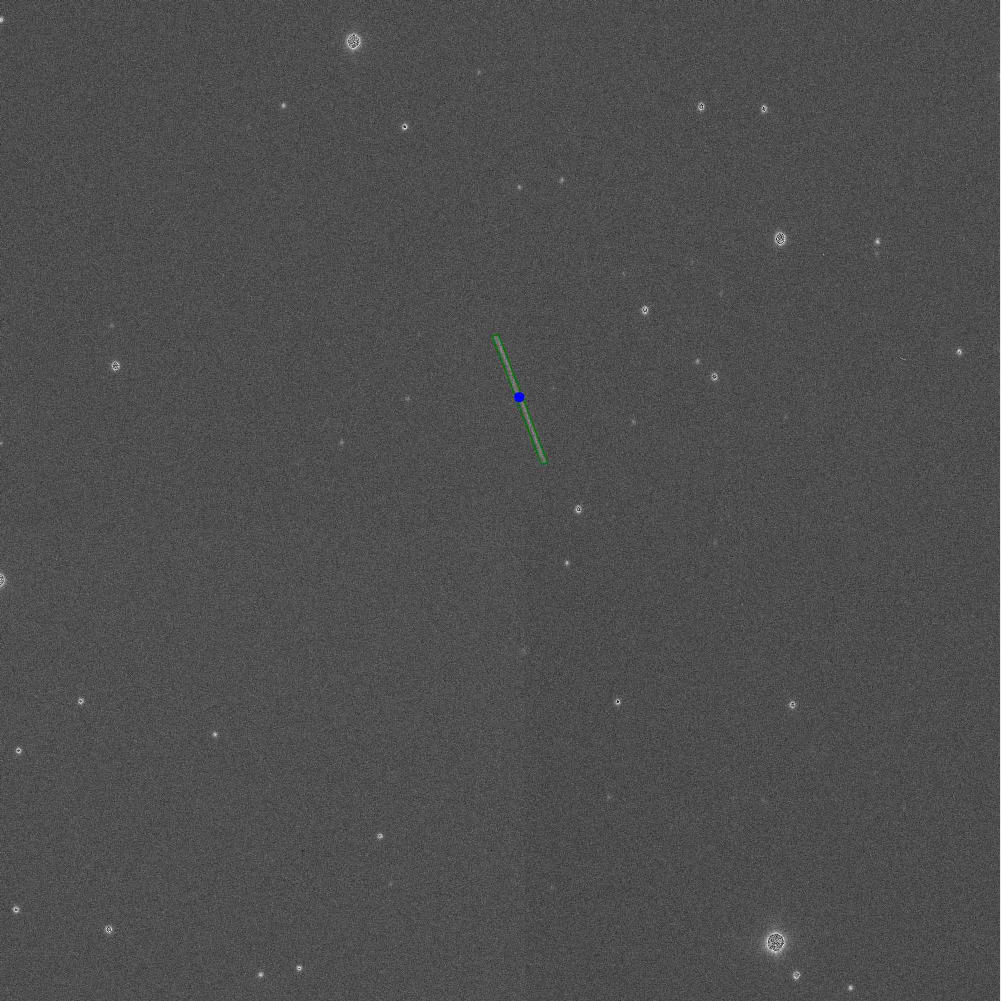}

\caption{Examples of detected satellite trails obtained during an observing campaign of Galileo satellites.}
\label{performance_examples}
\end{figure*}

The IAC80 SST system routinely produces a large number of measurements per night (see Figure \ref{performance_examples}). Under typical conditions, approximately 1400 measurements can be obtained per night for MEO and GEO objects, while fewer measurements are acquired for LEO targets due to their higher apparent motion.

The expected astrometric accuracy is better than 1 arcsecond across different orbital regimes. Detection efficiency is typically in the range of 80–90\% under nominal observing conditions.
The system has demonstrated stable performance over multiple years of operation, providing consistent and reliable data to the European SST network.
\section{Future work}
The IAC80 telescope has successfully evolved from a classical astronomical instrument into a fully operational asset for space debris tracking and surveillance. Through the development of dedicated instrumentation, automation tools, and a real-time data reduction pipeline, the telescope now contributes regularly to European SST activities.

Future developments include improvements in photometric calibration, enhanced satellite characterization using multi-filter observations, and further optimization of automation and scheduling capabilities. In addition, the IAC80 telescope will be upgraded soon with a modern control system, a wide-field camera and new engines to increase the angular tracking velocity from the current velocity of 0.5 degrees per second up to 1 degree per second.

\section{ACKNOWLEDGEMENTS}
This paper is based on observations performed with the IAC80 telescope of the Instituto de Astrofísica de Canarias at Observatorio del Teide.
The author would like to thank the IAC Telescope Operations Group and all collaborators involved in the development of CAMELOT2, CARONTE, and the SATRED pipeline. This work contributes to the European Space Surveillance and Tracking initiative.
The algorithms and codes reported here are available upon request to the corresponding author.
\renewcommand{\refname}{REFERENCES}
\bibliography{rmxac}

\begin{thebibliography}{}
\expandafter\ifx\csname natexlab\endcsname\relax\def\natexlab#1{#1}\fi
\providecommand{\url}[1]{\href{#1}{#1}}
\providecommand{\dodoi}[1]{doi:~\href{http://doi.org/#1}{\nolinkurl{#1}}}
\providecommand{\doeprint}[1]{\href{http://ascl.net/#1}{\nolinkurl{http://ascl.net/#1}}}
\providecommand{\doarXiv}[1]{\href{https://arxiv.org/abs/#1}{\nolinkurl{https://arxiv.org/abs/#1}}}

\bibitem[{{Barrientos} \& {Mendoza}(2017)}]{BarMen17}
{Barrientos}, E., \& {Mendoza}, S. 2017, European Physical Journal Plus, 132,
  361, \dodoi{10.1140/epjp/i2017-11642-2}

\bibitem[{{Feuersänger}(2012)}]{PFGPlots}
{Feuersänger}, C. 2012, {PGFPlots - A LaTeX package to create plots.}
\newblock \url{https://pgfplots.sourceforge.net/}

\bibitem[{{Rom{\'a}n-Z{\'u}{\~n}iga} {et~al.}(2015){Rom{\'a}n-Z{\'u}{\~n}iga},
  {Ybarra}, {Meg{\'\i}as}, {Tapia}, {Lada}, \& {Alves}}]{Roman+15}
{Rom{\'a}n-Z{\'u}{\~n}iga}, C.~G., {Ybarra}, J.~E., {Meg{\'\i}as}, G.~D.,
  {et~al.} 2015, \aj, 150, 80, \dodoi{10.1088/0004-6256/150/3/80}

\bibitem[{{Rom{\'a}n-Z{\'u}{\~n}iga} {et~al.}(2023){Rom{\'a}n-Z{\'u}{\~n}iga},
  {Kounkel}, {Hern{\'a}ndez}, {Pe{\~n}a Ram{\'\i}rez}, {L{\'o}pez-Valdivia},
  {Covey}, {Stutz}, {Roman-Lopes}, {Campbell}, {Khilfeh}, {Tapia},
  {Stringfellow}, {Downes}, {Stassun}, {Minniti}, {Bayo}, {Kim}, {Su{\'a}rez},
  {Ybarra}, {Fern{\'a}ndez-Trincado}, {Longa-Pe{\~n}a},
  {Ram{\'\i}rez-Preciado}, {Serna}, {Lane}, {Garc{\'\i}a-Hern{\'a}ndez},
  {Beaton}, {Bizyaev}, \& {Pan}}]{Roman23}
{Rom{\'a}n-Z{\'u}{\~n}iga}, C.~G., {Kounkel}, M., {Hern{\'a}ndez}, J., {et~al.}
  2023, \aj, 165, 51, \dodoi{10.3847/1538-3881/aca3a4}

\bibitem[{{Wikimedia projects contributors}(2023)}]{projects-2023}
{Wikimedia projects contributors}. 2023, {LaTeX/Tables},  {Pergamon Press}.
\newblock \url{https://en.wikibooks.org/wiki/LaTeX/Tables}

\end{thebibliography}


\begin{thebibliography}{}
\expandafter\ifx\csname natexlab\endcsname\relax\def\natexlab#1{#1}\fi
\providecommand{\url}[1]{\href{#1}{#1}}
\providecommand{\dodoi}[1]{doi:~\href{http://doi.org/#1}{\nolinkurl{#1}}}
\providecommand{\doeprint}[1]{\href{http://ascl.net/#1}{\nolinkurl{http://ascl.net/#1}}}
\providecommand{\doarXiv}[1]{\href{https://arxiv.org/abs/#1}{\nolinkurl{https://arxiv.org/abs/#1}}}

\bibitem[{{Instituto de Astrof\'isica de
  Canarias}(2025{\natexlab{a}})}]{camelot2}
{Instituto de Astrof\'isica de Canarias}. 2025{\natexlab{a}}, CAMELOT2
  Instrument,
  \url{https://research.iac.es/OOCC/iac-managed-telescopes/iac80/camelot2-2/}

\bibitem[{{Instituto de Astrof\'isica de
  Canarias}(2025{\natexlab{b}})}]{caronte}
---. 2025{\natexlab{b}}, CARONTE Instrument,
  \url{https://research.iac.es/OOCC/iac-managed-telescopes/iac80/caronte/}

\end{thebibliography}

\end{document}